\documentclass[twocolumn,showpacs,preprintnumbers,amsmath,amssymb,prb]{revtex4}

\usepackage{graphicx}
\usepackage{bm}

\begin{document}

\title{Characteristics of the Wave Function of Coupled Oscillators in Semiquantum Chaos}

\author{Gang Wu}
\email{wugaxp@gmail.com}
\author {Jinming Dong}
\email{jdong@nju.edu.cn}

\affiliation{Group of Computational Condensed Matter Physics,
National Laboratory of Solid State Microstructures and Department
of Physics, Nanjing University, Nanjing 210093, P. R. China}

\begin{abstract}

Using the method of adiabatic invariants and the Born-Oppenheimer
approximation, we have successfully got the excited-state wave functions for
a pair of coupled oscillators in the so-called \textit{semiquantum chaos}. Some interesting
characteristics in the \textit{Fourier spectra} of the wave functions and its \textit{Correlation Functions} in the regular and
chaos states have been found, which offers a new way to distinguish the
regular and chaotic states in quantum system.

\end{abstract}

\pacs {05.45. --a, 03.65.Sq}

\date{\today}
\maketitle

\section{ INTRODUCTION}

Even though plenty of research work has been done on chaos in quantum
system, the surprising fact is that its proper definition has not been
clearly got now.

From ``the principle of quantum corresponding'', there must be counterpart
of classical limit for a quantum dynamical system. Percival [1] showed that
the semiclassical approximation provides a way to divide the quantum energy
spectrum of a N-dimension conservative system into two parts, regular and
irregular, which relate, respectively, to periodic and chaotic motions in
their classical phase space. Bohigas \textit{et. al.} [2] and Berry \textit{et. al.} [3] then developed the
approach of energy-level statistics to describe the properties of regular
and irregular quantum energy spectra. Meanwhile, the studies of
eigenfunctions made by Shapiro [4] and Heller [5] give us more information
about the regular and irregular states of quantum systems.

In quantum mechanics, nonexistence of the concept of path or trajectory
prevents us from using the definition of classical chaos, which is
characterized by the sensitivity of the system to the initial conditions.
There are many theories to study properties of the irregular behavior, or
the so-called \textit{quantum chaos}, but most researchers have focused on the concept of
integrability in quantum mechanics and hope to get an exact definition of
the quantum chaos.

In this paper, we will consider a system in which a classical oscillator
interacts with a purely quantum-mechanical oscillator. This system was first
considered by Cooper \textit{et.al.} [6], which can be described by a \textit{classical effective Hamiltonian}, the expectation
value of the quantum Hamiltonian. Faccioli \textit{et.al.} in their comment [7] gave a more
proper treatment for this problem. But Cooper \textit{et.al. }and Faccioli \textit{et.al.} had only paid
their attention to the \textit{average value} of the time-dependent occupation number in the
chaotic state. As we all know, in quantum mechanics, the wave function can
give not only the static but also the dynamical information of the system.
Recently, Hui \textit{et. al.} in their paper [8] studied the ground-state wave function of
the coupled oscillators, and gave out a method to distinguish the chaotic
evolution in this system. But they did not study the behaviors in its
excited state. Then there come some questions: Will this method still work
in the excited-state? Could we get also the excited-state wave functions for
the coupled oscillators? What are the properties of the excited-state wave
functions?

In this paper, we will try to answer these questions, and then to find
whether there is another way to reach the goal. In section II, we will use
the method of adiabatic invariant [9] and the Born-Oppenheimer Approximation
[10-13] to obtain the wave functions in any excited-states. In section III,
we will use these wave functions to discuss the time evolution
characteristics of the system, and calculate their correlation functions.
Finally, we will end the article with some conclusions and discussions in
section IV.

\section{ EXCITED-STATE WAVE FUNCTIONS OF THE COUPLED HARMONIC OSCILLATORS}

In this paper, we mainly deal with a coupled system, in which a classical
oscillator interacts with a purely quantum mechanical oscillator, described
by a classical Lagrangian:

\begin{equation}
\label{eq1}
L = \frac{1}{2}\dot{x}^2 + \frac{1}{2}\dot{A}^2 -
\frac{1}{2}\left( {m^2 + e^2A^2} \right)x^2,
\end{equation}

\noindent
where the coordinates $x$ and $A$ describe, respectively, the motion of the
quantum oscillator and the classical one in the system. Using the
Born-Oppenheimer approximation, we can decouple the system into a classical
part and a quantum part, each of which can be handled easily.

From Lagrangian (1), a Schr\"{o}dinger equation for a state with
total energy $E_n $can be obtained:

\begin{equation}
\label{eq2}
\frac{1}{2}\left( { - \hbar ^2\frac{\partial ^2}{\partial x^2} + \omega
^2x^2 - \hbar ^2\frac{\partial ^2}{\partial A^2} - 2E_n } \right)\Psi _{E_n
} \left( {x,A} \right) = 0,
\end{equation}

\noindent
where $\omega ^2 = m^2 + e^2A^2$. Factorizing $\Psi _{E_n } \left( {x,A}
\right) = \psi \left( A \right)\chi _n \left( {x,A} \right)$ and following
the treatment in Ref. [7,8], $i.e.$, using the semiclassical approximation, we
obtain the following coupled equations for the $\psi \left( A \right)$ and
$\chi _n \left( {x,A} \right)$

\begin{equation}
\label{eq3}
\left( {\frac{1}{2}\dot {A}^2 + \left\langle {\hat {H}_x } \right\rangle _n
} \right)\psi = E_n \psi,
\end{equation}

\noindent
and

\begin{equation}
\label{eq4}
\left( {\hat {H}_x - i\hbar \frac{\partial }{\partial t}} \right)\chi _{sn}
= 0,
\end{equation}

\noindent
where

\begin{equation}
\label{eq5}
\hat {H}_x = \frac{1}{2}\left( { - \hbar ^2\frac{\partial ^2}{\partial x^2}
+ \omega ^2\hat {x}^2} \right),
\end{equation}

\begin{equation}
\label{eq6} \chi _n \left( {x,A} \right) = \\
\chi _{sn} \left( {x,A}
\right)\exp \left[ {\int ^{t}  {d{t}' \left( {\frac{i}{\hbar
}\left\langle {\hat {H}_x } \right\rangle _n + \left\langle
{\frac{\partial }{\partial {A}'}} \right\rangle _n {\dot {A}}'}
\right)} } \right],
\end{equation}

\noindent
and for an operator $\hat {O}$, its average value in $n_{th} $ excited state
is defined as

\begin{equation}
\label{eq7}
\left\langle \hat {O} \right\rangle _n = \frac{\int {dx\chi _n ^ * \hat
{O}\chi } _n }{\int {dx\chi _n ^ * \chi } _n }.
\end{equation}

Now, it is easy to solve the Schr\"{o}dinger equation, Eq. (\ref{eq4}), by using the
adiabatic invariant method mentioned in Ref.[1,7,8], in which an adiabatic
invariant $I(t)$ and the time-dependent canonical lowering and raising
operators $\hat {a}$ and $\hat {a}^\dag $ are introduced. In order to
compare with the paper of Cooper \textit{et.al.}, we let $G = \left\langle {x^2}
\right\rangle _0 $. Thus, $\hat {a}$ has the following form:

\begin{equation}
\label{eq8}
\hat {a} = e^{i\theta }\sqrt {G\hbar } \left[ {\frac{\partial }{\partial x}
+ \frac{1 - i\dot {G}}{2G\hbar }x} \right],
\end{equation}

\noindent
and

\begin{equation}
\label{eq9} \theta = \int ^t {dt'\frac{1}{2G}} .
\end{equation}

Further, when we have given the fixed total energy $E_n $, the $A$ and $G$
satisfy the following equations:

\begin{equation}
\label{eq10}
\ddot {A} = - \frac{\partial }{\partial A}\left\langle {\hat {H}_x }
\right\rangle _n = - \left( {2n + 1} \right)\hbar e^2AG,
\end{equation}

\begin{equation}
\label{eq11}
\frac{1}{2}\frac{\ddot {G}}{G} - \frac{1}{4}\left( {\frac{\dot {G}}{G}}
\right)^2 - \frac{1}{4G^2} + m^2 + e^2A^2 = 0.
\end{equation}

On defining the $n_{th} $ excited state of $a$ by $\textstyle{1 \over {\sqrt
{n!} }}\left( \hat {a} \right)^{n + 1}\left| n \right\rangle = 0$, we can
get in the coordinate representation,

\begin{equation}
\label{eq12}
\left( {\frac{\partial }{\partial x} + \frac{1 - i\dot {G}}{2G\hbar }x}
\right)^{n + 1}\phi _n \left( {x,G\left( t \right)} \right) = 0,
\end{equation}

\noindent
where $\phi _n \left( {x,G\left( t \right)} \right)$ is just the $n_{th} $
excited state wave function, i.e., $\left| n \right\rangle $. Its normalized
solution can be easily obtained as follows:

\begin{equation}
\label{eq13}
\left| n \right\rangle = \left( {\frac{1}{2G\hbar }} \right)^{\textstyle{1
\over 4}}H_n \left( {\frac{x}{\sqrt {2G\hbar } }} \right)\exp \left[ { -
\frac{1}{4G\hbar }\left( {1 - i\dot {G}} \right)x^2} \right],
\end{equation}

\noindent
where $H_n \left( x \right)$ is the $n_{th} $ Hermit function, defined as:
$H_n \left( x \right) = \left( { - 1} \right)^n\left( {2^nn!\sqrt \pi }
\right)^{ - \textstyle{1 \over 2}}\exp \left( {x^2}
\right)\frac{d^n}{dx^n}\exp \left( { - x^2} \right)$. And we take $\hbar =
1$ and $m = 1$ for simplicity.

From Eq. (\ref{eq13}), we have

\begin{equation}
\label{eq14}
\left| {\phi _n \left( {x,G\left( t \right)} \right)} \right|^2 =
\frac{1}{\sqrt {2G} }\left[ {H_n \left( {\frac{x}{\sqrt {2G} }} \right)}
\right]^2\exp \left( { - \frac{x^2}{2G}} \right).
\end{equation}

For a special case of $n = 0$, $\phi _0 \left( {x,G\left( t \right)} \right)
= \left( {\frac{1}{2\pi G}} \right)^{\frac{1}{4}}\exp \left[ { - \left(
{\frac{1 - i\dot {G}}{2G}} \right)\frac{x^2}{2}} \right]$, which is just the
ground state wave functions given in Ref. [8].

And it is easy to get the total energy and the quantum part energy:

\begin{equation}
\label{eq15}
E_n = \frac{1}{2}\dot{A}^2 + \left( {2n + 1}
\right)\left[ {\frac{\dot{G}^2 + 1}{8G} + \frac{1}{2}\omega ^2G}
\right],
\end{equation}

\begin{equation}
\label{eq16}
E_{Qn} = \left( {2n + 1} \right)\left[ {\frac{\dot{G}^2 + 1}{8G}
+ \frac{1}{2}\omega ^2G} \right] = E_n - \frac{1}{2}\dot{A}^2 \le E_n .
\end{equation}

Especially, when $n = 0$, $E_0 = \frac{1}{2}\dot{A}^2
+ \frac{\dot{G}^2 + 1}{8G} + \frac{1}{2}\omega ^2G =
\frac{1}{2}\dot{A}^2 + E_{Q0}$. 

In order to compare with the results in Ref. [8], we will only fix the total system energy
(then, the initial condition and the phase space are fixed.) in the
following discussion.

\section{DISCUSSIONS ABOUT THE EXCITED-STATE WAVE FUNCTION}

\begin{figure}[htbp]
\includegraphics[width=\columnwidth]{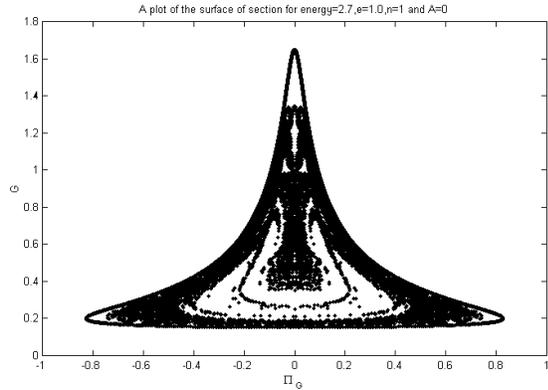}
\caption{The Poincar\'{e} section in the phase space.}
\label{fig1}
\end{figure}

Utilizing Eq. (\ref{eq13}), we can obtain the \textit{Fourier spectra} (FS) of probability density operator
in the excited-state wave functions in chaotic or regular state at the fixed
position $x$. And from them, we can see whether the phenomena happened in
ground-state case would be possible to appear in excited-state. For
simplicity, we will use the units of $\hbar = 1$ and $m = 1$, and we take
the first excited-state for example. First we show the Poincar\'{e} section
in the phase space in Fig. 1 by solving the coupled equations (9.5a) and
(9.5b).

From Fig. 1, we can easily select the initial conditions as the following:

¢Å Chaos State: $A = 0.0,\dot {A} = 1.48565,G = 0.35,\dot {G} = 0.0;e =
1.0;$

¢Æ Regular State: $A = 0.0,\dot {A} = 1.17969,G = 0.225,\dot {G} = 0.0;e =
1.0;$

And we calculated their \textit{Lyapunov Exponents} (LE) to verify our selection. We find the LE in the
first condition is positive and the second one is zero. Fig. 2 shows the FS
of the probability density in the first excited state ($n = 1)$.

\begin{figure}[htbp]
\includegraphics[width=\columnwidth]{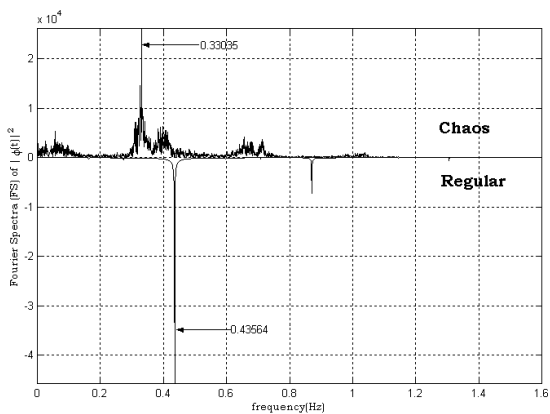}
\caption{Fourier spectra of the time evolution of the probability,
$\left| {\phi _n \left( {x,G\left( t \right)} \right)} \right|^2$, where
n=1.}
\label{fig2}
\end{figure}

From Fig. 2., we can see that in the first excited-state, those
\textit{`fundamental frequencies' }still can be found in the Fourier spectra, such as $\nu _1^R $=0.4356 Hz,$_{
}\nu _1^C $=0.3304 Hz,\textit{ etc}. They still play important roles in combining the
frequencies in the FS.

But if we compare the FS in the first-excited state and the ground states,
we can find much difference between them.

(a). It's obvious that in the first-excited state the FS in regular state
appears more regular than the one in ground state, while the FS of chaotic
state appears more chaotic than the one in the ground state.

(b). We can find that in the first-excited state the fundamental frequencies
in regular state are larger than the corresponding ones in the ground state,
while the fundamental frequencies in chaotic state are smaller than the
corresponding ones in the ground state.

In fact, these phenomena can be explained as follows:

(a). It is pointed ([8]) that the \textit{`fundamental frequencies' }originate only from the coupling between
the classical and quantum parts. Because of that reason, the \textit{`fundamental frequencies' }will also
appear in the excited states.

(b). In the first-excited state, the LE of chaotic state ($G\left( 0 \right)
= 0.35)$ is almost two times larger than that in the ground state ($G\left(
0 \right) = 0.5)$. And we find the LE in regular state ($G\left( 0 \right) =
0.225)$ converges to zero faster than that in the ground state ($G\left( 0
\right) = 0.35)$. So in excited-state, the chaotic state will become more
chaotic, and regular state will appear more regular.

From above discussion, we can get a conclusion that the difference between
regular and chaotic state in FS still clearly exists in the first excited
state. Because of that reason we can use the FS to distinguish the chaotic
and regular states even in excited-states, too.

Now, we study the correlation function in the excited states. The
\textit{Correlation Function} (CF) is defined as

\begin{equation}
\label{eq17}
CF\left( t \right) = \mathop {\lim }\limits_{L \to \infty }
\frac{1}{2L}\int\limits_{ - L}^{ + L} {dt' \\
\int\limits_{ - \infty }^{ +
\infty } {dx\left| {\phi \left( {x,t + t'} \right)} \right|\left| {\phi
\left( {x,t'} \right)} \right|} },
\end{equation}

\begin{figure}[htbp]
\includegraphics[width=\columnwidth]{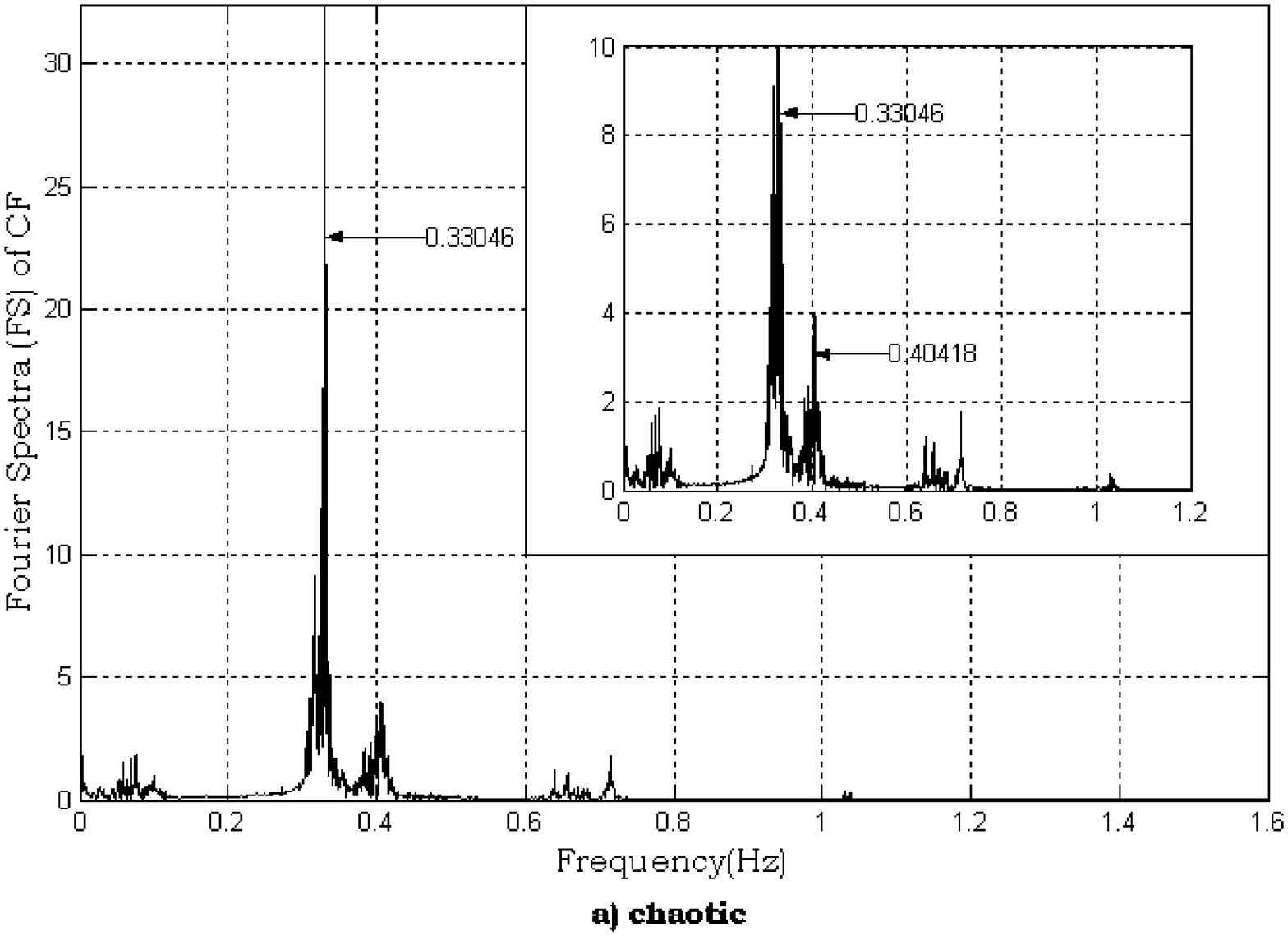}
\includegraphics[width=\columnwidth]{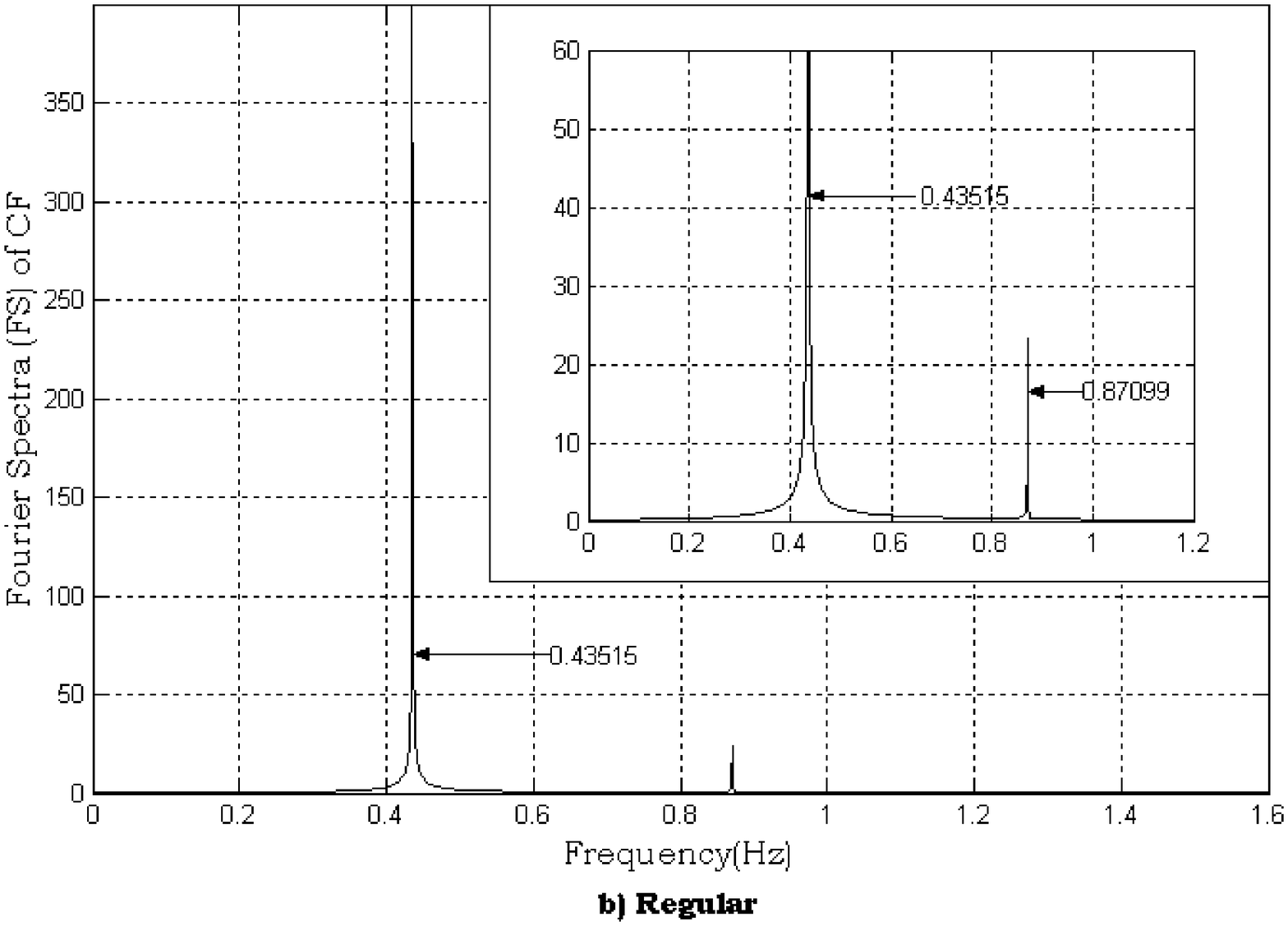}
\caption{Fourier spectra of the time-correlation functions of
$\left| {\phi _n \left( {x,G\left( t \right)} \right)} \right|^2$, n=1. a)
in chaotic state; b) in regular state.}
\label{fig3}
\end{figure}

\begin{figure}[htbp]
\includegraphics[width=\columnwidth]{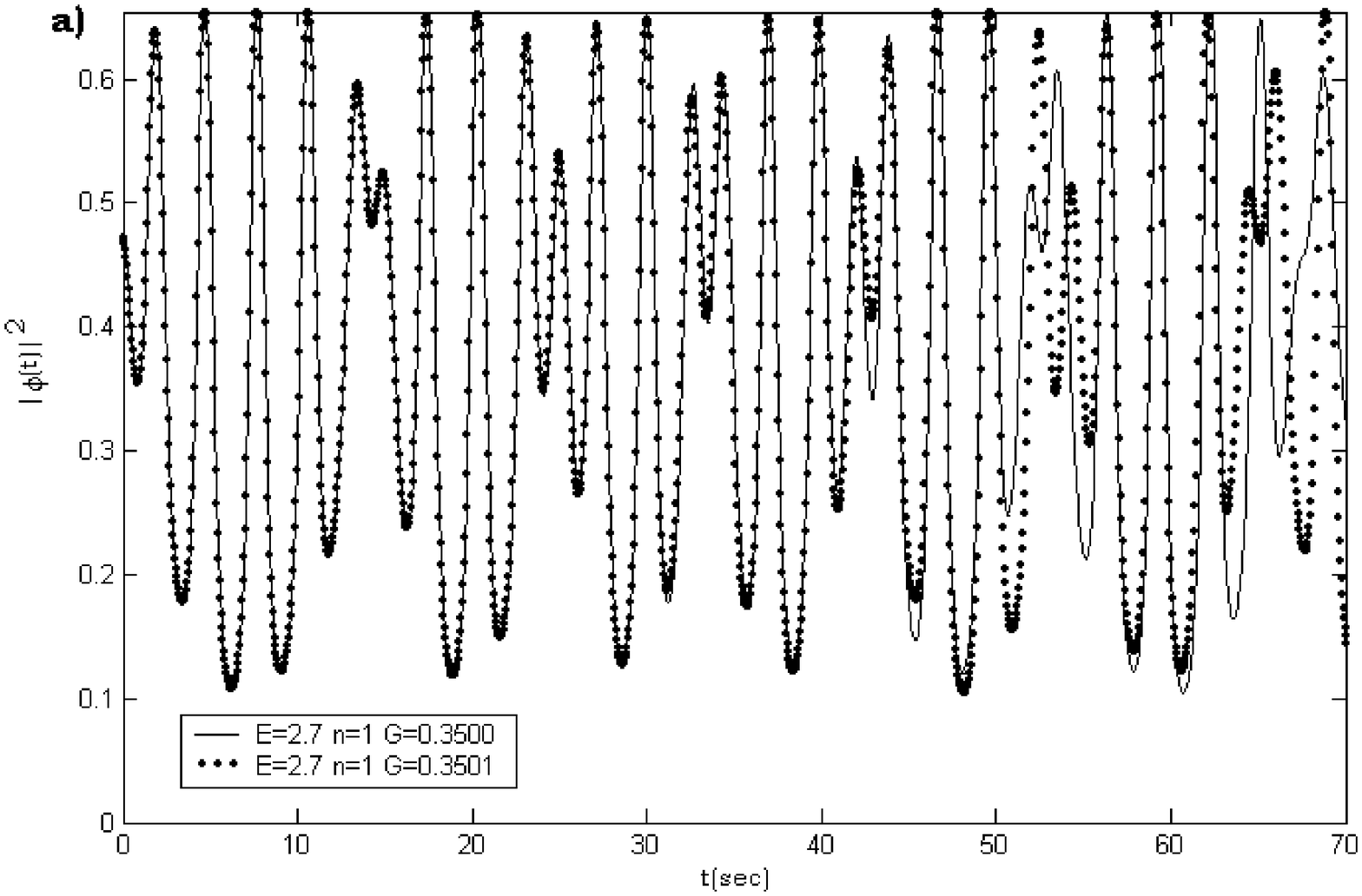}
\includegraphics[width=\columnwidth]{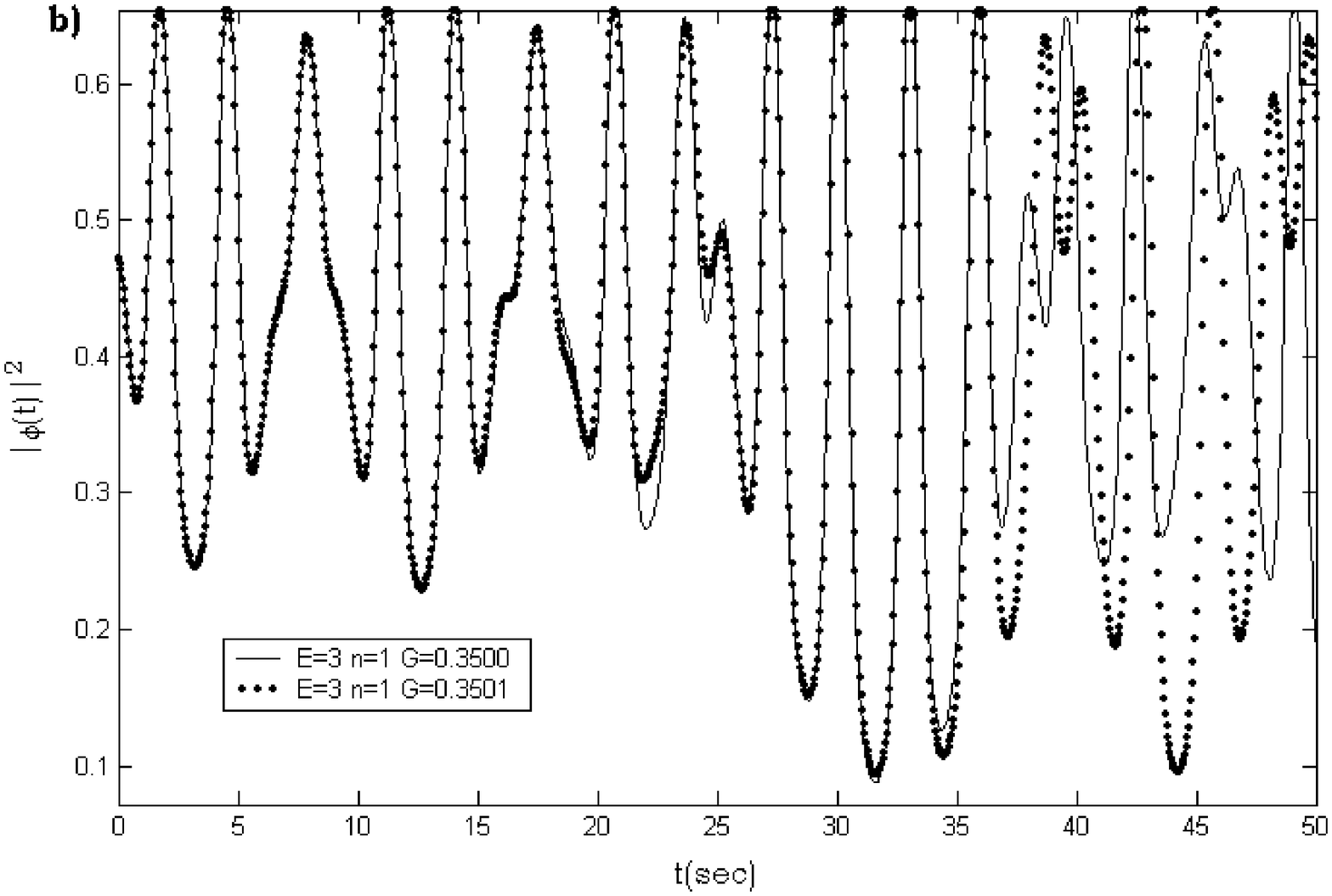}
\caption{Sensitivity of the probability density to the initial
conditions for different energies. a) $E_1 = 2.7$; b) $E_1 = 3.0$.}
\label{fig4}
\end{figure}

We have got the excited wave function (Eq. 11), so we can get the
time-correlation functions in the chaotic and regular state and their FS,
respectively.

\[
CF\left( t \right) = \mathop {\lim }\limits_{L \to \infty }
\frac{1}{L}\int_{ - L}^{ + L} {dt'} \int_{ - \infty }^{ + \infty } {dx\left[
{4G\left( {t'} \right)G\left( {t + t'} \right)} \right]^{ - \frac{1}{4}}}
\]

\[
\times H_n \left( {\frac{x}{\sqrt {2G\left( {t'} \right)} }} \right)H_n \left(
{\frac{x}{\sqrt {2G\left( {t + t'} \right)} }} \right)
\]

\[
\times \exp \left\{ { - \frac{x^2}{4}\left[ {\frac{1}{G\left( {t'} \right)}
+ \frac{1}{G\left( {t + t'} \right)}} \right]} \right\}
\]

\[
 = \mathop {\lim }\limits_{L \to \infty } \frac{1}{2L}\int_{ - L}^{ + L}
{dt'\frac{\sqrt c }{2^nn!}\left\{ {\left[ {c_1 \frac{k^n - 1}{k - 1} + c_1 ^
* \frac{\left( {k^ * } \right)^n - 1}{\left( {k^ * } \right) - 1}} \right] \\
+ \left[ {\frac{1}{2}\left( {1 + c^2} \right)} \right]^{ - \frac{1}{2}}}
\right\}}
\]

\[
 = \mathop {\lim }\limits_{L \to \infty } \frac{1}{2L}\int_{ - L}^{ + L}
{dt'\frac{\sqrt c }{2^nn!}\left\{ {2Re\left[ {c_1 \frac{k^n - 1}{k - 1}}
\right] + \left[ {\frac{1}{2}\left( {1 + c^2} \right)} \right]^{ -
\frac{1}{2}}} \right\}},
\]

\noindent
where

\[
c_1 = \alpha _1 + i\beta _1 ,k = \alpha _2 + i\beta _2,
\]

\noindent
and

\[
\alpha _1 = - \frac{1}{2}\frac{c^2 - 4c + 1}{\left( {1 + c^2} \right)^{3 /
2}},
\]

\[
\beta _1 = \sqrt 2 \left( {\frac{1}{1 + c^2}}
\right)^{\frac{3}{2}}\frac{c^4 - 6c^2 + 1}{c^2 - 1},
\]

\[
\alpha _2 = \frac{4c}{1 + c^2},\beta _2 = \frac{2\left( {c^2 - 1} \right)}{1
+ c^2},
\]

\[
c = \sqrt {\frac{G\left( {t'} \right)}{G\left( {t + t'} \right)}} _{.}
\]

In Fig. 3, we give out the Fourier spectra of time-correlation function in
the first excited state, in which the mean value has been subtracted from
the FS to remove the big dc-component in the FS. In these figures, we can
find that those \textit{`fundamental frequencies' }appear again, and these figures can show the \textit{`fundamental frequencies' }more clearly.
So we know that, the FS of time-correlation can be a justifiable way to
distinguish the regular and chaotic states. And when the system is in the
chaotic state, the FS of the correlation function becomes desultory, and it
looks to have a lot of `noise' components.

Finally, we examine the sensitivity of the eigenfunctions in the chaotic
state to the initial conditions, which is usually considered as a basic
characteristic of chaos.

Fig.4 shows related numerical results of the probability density $\left|
{\phi _n \left( {x,G\left( t \right)} \right)} \right|^2$ \textit{vs.} t for the case of
$E_1 = 2.7$ (Fig. 4a) and 3.0 (Fig. 4b). The initial conditions are selected
by following rule: all parameters are the same except that $G(0)$ has very
little difference between two solutions, which are $G(0) = 0.35$ and $G(0) =
0.3501$. We can see that in the first excited state, there still exists such
sensitivity for the eigenfunction, which is influenced by the total system
energy $E_1 $. And according to Ref. [8], when the system energy rises, the
chaotic behavior becomes stronger.

\section{CONCLUSIONS}

In this paper, we analytically calculated the excited-state wave functions
of a quantum oscillator coupled with a classical harmonic oscillator.
Instead of using the definition of the classical chaos, such as the
sensitivity of the system to the initial conditions, we have found the
\textit{`fundamental frequencies'} in the ground state still appear in the excited state, and found a very
useful way to get these characteristics. We think these characteristics can
be used to distinguish the regular and chaotic states when the so-called
\textit{semiquantum chaos} emerges in the actual system and our result is helpful to further research
work in this field.

\end{document}